\documentclass[twocolumn]{aastex631}

\hypersetup{linkcolor=red,citecolor=blue,filecolor=cyan,urlcolor=magenta}

\usepackage{graphicx}
\newif\iffigs
\newif\iffigscl
\newif\iffigstest
\newif\iflabs

\figstrue

\figscltrue

\figstestfalse

\labsfalse

\shorttitle{Morphological estimate of the age of open clusters}
\shortauthors{Dinnbier et al.}

\graphicspath{{./}{figures/}}

\begin{document} 

\newif\iffigs

\figstrue

\newcommand{\itii}[1]{{#1}}
\newcommand{\franta}[1]{\textbf{\color{green} #1}}
\newcommand{\frantaii}[1]{\textbf{\color{yellow} #1}}
\newcommand{\itiitext}[1]{{#1}}

\newcommand{\eq}[1]{eq. (\ref{#1})}
\newcommand{\eqp}[1]{(eq. \ref{#1})}
\newcommand{\eqq}[1]{eq. \ref{#1}}
\newcommand{\eqb}[2]{eq. (\ref{#1}) and eq. (\ref{#2})}
\newcommand{\eqc}[3]{eq. (\ref{#1}), eq. (\ref{#2}) and eq. (\ref{#3})}
\newcommand{\refs}[1]{Sect. \ref{#1}}
\newcommand{\reff}[1]{Fig. \ref{#1}}
\newcommand{\reft}[1]{Table \ref{#1}}

\newcommand{\datum} [1] { \noindent \\#1: \\}
\newcommand{\pol}[1]{\vspace{2mm} \noindent \\ \textbf{#1} \\}
\newcommand{\code}[1]{\texttt{#1}}
\newcommand{\figpan}[1]{{\sc {#1}}}

\newcommand{\nbdvi}{\textsc{nbody6} }
\newcommand{\nbdvid}{\textsc{nbody6}}
\newcommand{\flash}{\textsc{flash} }
\newcommand{\flashd}{\textsc{flash}}
\newcommand{\sfe}{\mathrm{SFE}}
\newcommand{\mum}{$\; \mu \mathrm{m} \;$}
\newcommand{\rop}{$\rho$ Oph }
\newcommand{\HT}{$\mathrm{H}_2$}
\newcommand{\Halpha}{$\mathrm{H}\alpha \;$}
\newcommand{\HI}{H {\sc i} }
\newcommand{\HII}{H {\sc ii} }
\renewcommand{\deg}{$^\circ$}

\newcommand{\dd}{\mathrm{d}}
\newcommand{\acosh}{\mathrm{acosh}}
\newcommand{\sign}{\mathrm{sign}}
\newcommand{\cex}{\mathbf{e}_{x}}
\newcommand{\cey}{\mathbf{e}_{y}}
\newcommand{\cez}{\mathbf{e}_{z}}
\newcommand{\cer}{\mathbf{e}_{r}}
\newcommand{\ceR}{\mathbf{e}_{R}}

\newcommand{\llg}[1]{\log_{10}#1}
\newcommand{\pder}[2]{\frac{\partial #1}{\partial #2}}
\newcommand{\pderrow}[2]{\partial #1/\partial #2}
\newcommand{\nder}[2]{\frac{\dd #1}{\dd #2}}
\newcommand{\nderrow}[2]{{\dd #1}/{\dd #2}}

\newcommand{\Cmiii}{\, \mathrm{cm}^{-3}}
\newcommand{\Gcmii}{\, \mathrm{g} \, \, \mathrm{cm}^{-2}}
\newcommand{\Gcmiii}{\, \mathrm{g} \, \, \mathrm{cm}^{-3}}
\newcommand{\Kms}{\, \mathrm{km} \, \, \mathrm{s}^{-1}}
\newcommand{\Si}{\, \mathrm{s}^{-1}}
\newcommand{\Esi}{\, \mathrm{erg} \, \, \mathrm{s}^{-1}}
\newcommand{\Ee}{\, \mathrm{erg}}
\newcommand{\Yr}{\, \mathrm{yr}}
\newcommand{\Kyr}{\, \mathrm{kyr}}
\newcommand{\Myr}{\, \mathrm{Myr}}
\newcommand{\Gyr}{\, \mathrm{Gyr}}
\newcommand{\Msun}{\, \mathrm{M}_{\odot}}
\newcommand{\Rsun}{\, \mathrm{R}_{\odot}}
\newcommand{\Pc}{\, \mathrm{pc}}
\newcommand{\Kpc}{\, \mathrm{kpc}}
\newcommand{\Mpc}{\, \mathrm{Mpc}}
\newcommand{\Sd}{\Msun \, \Pc^{-2}}
\newcommand{\Ev}{\, \mathrm{eV}}
\newcommand{\Kk}{\, \mathrm{K}}
\newcommand{\Au}{\, \mathrm{AU}}
\newcommand{\Mas}{\, \mu \mathrm{as}}

\title{Estimating the ages of open star clusters from properties of their extended tidal tails}


\correspondingauthor{Franti\v{s}ek Dinnbier}
\email{dinnbier@sirrah.troja.mff.cuni.cz}

\author[0000-0001-5532-4211]{Franti\v{s}ek Dinnbier}
\affiliation{Astronomical Institute, Faculty of Mathematics and Physics, Charles University in Prague, V Hole\v{s}ovi\v{c}k\'{a}ch 2, 180 00 Praha 8, Czech Republic}

\author[0000-0002-7301-3377]{Pavel Kroupa}
\affiliation{Astronomical Institute, Faculty of Mathematics and Physics, Charles University in Prague, V Hole\v{s}ovi\v{c}k\'{a}ch 2, 180 00 Praha 8, Czech Republic}
\affiliation{Helmholtz-Institut f\"{u}r Strahlen- und Kernphysik, University of Bonn, Nussallee 14-16, 53115 Bonn, Germany}
\email{pkroupa@uni-bonn.de}

\author[0000-0003-1924-8834]{Ladislav \v{S}ubr}
\affiliation{Astronomical Institute, Faculty of Mathematics and Physics, Charles University in Prague, V Hole\v{s}ovi\v{c}k\'{a}ch 2, 180 00 Praha 8, Czech Republic}

\author[0000-0002-1251-9905]{Tereza Je\v{r}\'{a}bkov\'{a}}
\affiliation{European Space Agency (ESA), European Space Research and Technology Centre (ESTEC), Keplerlaan 1, 2201 AZ Noordwijk, The Netherlands}

\begin{abstract}

The most accurate current methods for determining the ages of open star clusters, stellar associations and stellar
streams are based on isochrone fitting or the lithium depletion boundary.
We propose another method for dating these objects based on the morphology of their extended tidal tails, which have been recently discovered
around several open star clusters.
Assuming that the early-appearing tidal tails, the so called tidal tails I, originate from the stars released from the cluster during early gas expulsion,
or that they form in the same 
star forming region as the cluster (i.e. being coeval with the cluster),
we derive the analytical formula for the tilt angle $\beta$ between the long axis of the tidal tail and the orbital direction
for clusters or streams on circular trajectories.
Since at a given Galactocentric radius, $\beta$ is only a function of age $t$ regardless of the initial properties of the cluster,
we estimate the cluster age by inverting the analytical formula $\beta = \beta (t)$.
We illustrate the method on a sample of $12$ objects, which we compiled from the literature, and we find a reasonable agreement with previous
dating methods in $\approx 70$\% of the cases.
This can probably be improved by taking into account the eccentricity of the orbits and by revisiting the dating methods based on stellar evolution.
The proposed morphological method is suitable for relatively young clusters (age $\lesssim 300 \Myr$),
where it provides a relative age error of the order of $10$ to $20$\%
for an error in the observed tilt angle of $5 ^\circ$.

\end{abstract}



%

\section{Introduction}

\label{sIntro}


Together with stellar mass and chemical composition, stellar age is one of the most important properties characterising a star,
and methods of its determination are of great interest. 
The knowledge of the stellar age is important, for example, for testing theories of stellar evolution, for reconstructing the star 
formation history of the Galaxy, and the relationship between 
enrichment and kinematics in the Galaxy, and for estimating the life-times of some circumstellar structures (e.g. debris discs). 

Two most accurate methods for determining the ages of open star clusters are isochrone fitting and the lithium depletion boundary. 
However, each method suffers from limitations. 
Isochrone fitting is sensitive to not fully understood processes in stellar evolution such as convective overshooting, rotational mixing, 
internal gravity waves and diffusion \citep{Meynet2009,Soderblom2010}. 
Moreover, stellar rotation prolongs stellar life-times and increases the luminosity for a given stellar mass \citep{Meynet2000}. 
Blue stragglers, if present, add another bias \citep{Brandt2015} as they appear as stars of a younger age. 
These uncertainties in stellar evolution introduce an error in the age determination of the order of tens of a percent \citep{Meynet2009}. 
Systematic errors (e.g. omission of the most luminous stars) can easily lead to much larger errors.  
For example, \citet{Meingast2019} overestimate the age of the Pisces-Eridanus stream by a factor of $8$ as later corrected by \citet{Curtis2019}. 

The other dating method, based on the lithium depletion boundary, is considered to be more accurate (but see \citealt{Jeffries2017,Bouvier2018} for complications 
due to rotation and uncertainties in stellar radii).
This method can be applied only for the most nearby clusters because of the very low luminosity of the relevant objects. 
Up to date, only a dozen clusters have ages determined by this method. 
In general, the lithium depletion boundary method provides older ages (by $\approx 50$\%) than isochrone fitting \citep[e.g.][]{Soderblom2010,Binks2021}. 
The difference between these two dating methods is not restricted to the youngest clusters. 
For example, \citet{Stauffer1998} estimate the age of the Pleiades based on lithium depletion boundary to be $\approx 125 \Myr$, 
while isochrone fitting provides the age from $70$ to $100 \Myr$.

Another reason for dating clusters is that they can be used as anchors for other (and generally less accurate) dating methods, 
for example gyrochronology, chromospheric emission, lithium abundance 
and coronal X-ray emission \citep[e.g.][]{Skumanich1972,Soderblom1991,Lachaume1999,Barnes2007,Mamajek2008}, 
which are instrumental in estimating the ages of field stars. 
Unlike isochrone fitting and the lithium depletion boundary, these methods lack a theoretical background, and they need to be calibrated empirically on 
coeval stellar systems such as star clusters or associations.

In addition to the dating methods described above, which are based on stellar evolution, there are also attempts to provide age estimates 
from stellar kinematics or cluster dynamics. 
Kinematical methods usually backtrace a population of stars until they assume the smallest volume, which is taken as the time of formation, 
or they estimate the onset of expansion from the current size and expansion velocity. 
Kinematical estimates are usually less accurate than those using stellar evolution, and they can be used only for the youngest objects (age $\lesssim 20 \Myr$) 
because of gravitational interactions within the cluster and the external field of the Galaxy, which bends the stellar trajectories \citep{Blaauw1991,Brown1997}. 
More recently, \citet{Crundall2019} suggest a more sophisticated tool taking into account the external Galactic field as well as the 
complicated morphology of star forming regions, extending the usability of the kinematic method up to $\approx 100 \Myr$ for objects with a 
sufficiently low velocity dispersion. 
Dynamical methods compare numerical models of star clusters at different degree of dynamical evolution with observations. 
They compare the radial mass distribution \citep{Buchholz1980}, or the ratio of the number of stars in two different mass bins \citep{Kroupa1995c}.
Dynamical methods are rarely used.

In this work, we present another method, which utilises stellar kinematics but in a very different way than the previous methods. 
Instead of tracing individual stars backward or forward in time, we investigate the shape of the extended tidal structure which is formed from 
an expanding group of stars. 
The expanding stellar structure is supposed to be formed of stars which escape from a star cluster 
as the result of gas expulsion \citep[e.g.][]{Lada1984,Kroupa2001b,Geyer2001}, which terminates its embedded phase.
The remnant of the star cluster which survives this event is accompanied by these stars. 
The existence of the extended stellar structures was predicted theoretically from the results of N-body simulations \citep{Kroupa2001b}, 
and further explored by \citet{Dinnbier2020a}, 
before similar structures surrounding star clusters were observed \citep{Meingast2021} thanks to the 
unprecedented sensitivity of the Gaia mission \citep{Gaiac2016b,Gaiac2018}.
At least some of these structures are coeval with the star cluster \citep{Bouma2021}, which supports the view of their common origin.
In an alternative scenario where the extended structures were never bound to the clusters, but probably formed close to them and are coeval with them, 
this dating method can be applied as well.

We introduce the new method for age determination in \refs{sMethodDescr}, and then apply it to 
the open star clusters which are surrounded by known extended structures in \refs{sApplication}.
In addition, we analyse two gravitationally unbound stellar streams. 
The method is discussed in \refs{sDiscussion}. 
We conclude in \refs{sSummary}.

\section{The time evolution of the tilt and the size of the tidal tail}

\label{sMethodDescr}

\iffigs
\begin{figure*}
\includegraphics[width=\textwidth]{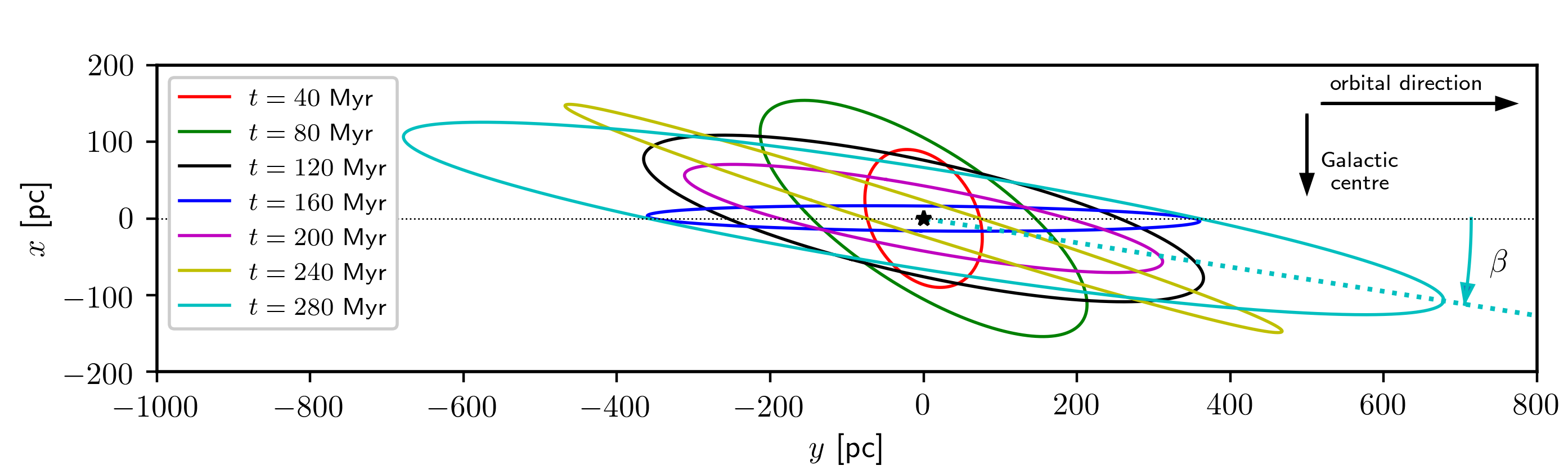}
\caption{
The orientation and shape of tidal tail I as calculated according to eqs. \ref{ePosition} and \ref{eXasAlpha} for 
stars escaping at $\widetilde{v}_{\rm e,I} = 2 \Kms$, and for the conditions at the solar circle.
The age of the tail is indicated by the colour. 
The star cluster is located at the centre of the coordinate system (the black star), and it orbits the Galaxy in the direction indicated 
at upper right. 
As the cluster and tail age, the direction of the long axis of the tidal tail changes from pointing almost towards the Galactic centre (at $t = 40 \Myr$), 
to the direction of the cluster motion (at $t = 160 \Myr$), and with increasing tilt again afterwards. 
Also note that the shape and aspect ratio of the tidal tail undergoes complicated changes with time.
The dotted cyan line shows the definition of the tail tilt angle $\beta$.
}
\label{ftailShape}
\end{figure*} \else \fi

We assume that star clusters form with a relatively low star formation efficiency ($\sfe \approx$ stellar mass/stellar and gaseous mass within the star forming volume)
\citep{Lada2003,Megeath2016,Banerjee2018}, 
and that they expel the non-star forming gas on a time-scale which is short in comparison to the cluster crossing time. 
These conditions unbind a substantial fraction of stars from the cluster (typically more than 60\%; \citealt{Lada1984,Goodwin1997,Kroupa2001b,Baumgardt2007}), 
which expand and form an extended tidal structure surrounding the cluster \citep{Dinnbier2020a, Dinnbier2020b} (hereafter Papers I and II, respectively). 
We refer to this tidal structure as \textit{tidal tail I}
\footnote{Although extended structures surrounding star clusters have been referred to by various terms in the literature (e.g. "halos", "coronae", "strings"; 
\citealt{Bouma2021}), we use the term tidal tail in the present work because we assume that they originate as tidal structures related to the cluster.}
.
The post-gas expulsion cluster revirialises \citep{Banerjee2013}, and loses stars gradually due to encounters between stars,
producing the classical S-shaped tidal tail \citep{Chumak2006a,Kupper2008,Kupper2010}, which is refereed to as \textit{tidal tail II}.
On the time-scale of several hundreds of Myr, tail I contains substantially more stars and is more extended than tail II \citep{Dinnbier2020b}.
The tidal tails of type I and II have been recently explored using observational data \citep[e.g.][]{Pang2021}.

The method takes advantage from the following kinematic property of stars. 
Stars which escape from the cluster through gas expulsion have the Galactic orbital velocity either 
slightly larger or smaller than the orbital velocity of the cluster. 
The stars with larger orbital velocity have the guiding centres of their orbits outside the orbit of the cluster (which is at Galactocentric radius $R_{\rm 0}$), 
while the stars with smaller orbital velocity have the guiding centres inside the orbit of the cluster. 
The former stars trail behind the cluster, while the latter overtake it. 
Consequently, the volume occupied by the escaping stars gradually stretches from a sphere to an elongated spheroid, 
and the direction $\beta$ (see its meaning in \reff{ftailShape}) of the long axis of the spheroid gets more aligned with the direction of the cluster orbit with time. 
Since all the stars escaped almost at the same time and they follow epicyclic motions with the same epicyclic frequency $\kappa$ to a good approximation, 
they reach their birth Galactocentric radius $R_{\rm 0}$ at the same time at $2\pi n/\kappa$, where $n$ is a positive integer,
but the stars are stretched along the azimuthal direction. 
This means that the tidal tail is aligned with the direction of motion each time $2\pi n/\kappa$, and it tilts relative to this 
direction in between these time events.
Thus, the age of the tail can be estimated by inverting the theoretical time dependence of the tail tilt $\beta = \beta(t)$.

The width of the stellar structure depends on the characteristic speed of escaping stars, which is a function of the embedded cluster mass, 
and the phase of the tail oscillation (for example, the thickness of the idealised tail drops to zero at $2\pi n/\kappa$). 
The width of the tail at the known age can constrain the initial mass of the cluster. 
In this section, we derive the time dependence of two quantities: the tilt (\refs{ssDynAge}) and width (\refs{ssMass}) of the tidal tail.

The present semi-analytic study is an extension of the analysis of Paper I, it uses the same assumptions, and it applies only to tail I.
The first assumption is that star clusters release many stars during a time window whose duration is short in comparison to the epicyclic 
time scale $2\pi/\kappa$ (which is $\approx 168 \Myr$ at the Solar circle, \citealt{Allen1991}). 
The physical process which is responsible for unbinding the stars is assumed to be gas expulsion of the residual gas from the newly formed open star clusters; 
however the solution is general, and it applies to any isotropically expanding stellar system in an external tidal field as 
long as the initial size of the stellar system is significantly smaller than the Galactocentric distance. 
The more general examples include a shock caused by an encounter between a star cluster (not necessarily young) and a molecular cloud, 
or a dissolution of a population of sparse clusters as their natal clouds are disrupted, forming a gravitationally unbound stellar stream.

The second assumption is that the speeds of escaping stars $\widetilde{v}_{\rm e,I}$ are approximately 
equal or larger than the speed $\widetilde{v}_{\rm e,II}$ corresponding to 
the difference between the maximum and minimum of the gravitational potential around the tidal radius $r_{\rm t}$. 
This condition means that stars of typical velocity $\widetilde{v}_{\rm e,I}$ can overcome the Jacobi potential at any direction without 
significant change to $\widetilde{v}_{\rm e,I}$, and thus escape the cluster with comparable probability in any direction. 
As shown in paper I (sect. 2.1 and 4.7 there), this condition is fulfilled for practically all embedded star clusters currently forming in the Galaxy. 

We adopt the usual coordinate system, where the $x$-axis points in the direction of the Galactic anti-centre, and the $y$-axis points in the direction of the 
Galactic rotation
\footnote{The adopted coordinate system is left-handed because the Galaxy rotates clockwise when seen from the North Galactic Pole \citep{Binney2008}. 
This choice, which is more convenient for comparison with observations, is different from Papers I and II, where a right-handed system was adopted.}
. 
The star cluster is on a circular orbit, so it is at rest at the origin of the coordinate axes. 
In this idealised model, we assume that all stars escape the cluster with the same velocity $\widetilde{v}_{\rm e,I}$. 
The components of the velocity vector ($v_{\rm R}$, $v_{\rm \phi}$), where $\widetilde{v}_{\rm e,I}^2 = v_{\rm R}^2 + v_{\rm \phi}^2$, 
are parallel to the coordinates ($x$, $y$), which lie in the plane of the Galaxy. 
The vertical motion is neglected. 

The star escapes the cluster at an angle $\alpha$, whose meaning is shown in fig. 1 of paper I, and then moves along an epicycle 
of a guiding centre at Galactocentric radius $R_{\rm g}$, and of semi-major axis $Y$, and semi-minor axis $X$. 
The guiding centre itself moves in the azimuthal direction at velocity $v_{\rm g}$ relative to the cluster. 
The relative position of the star to the cluster at time $t$ (since its escape) is given by (c.f. eq. 12 in Paper I)
\begin{eqnarray}
x(\alpha, t) = && X(\alpha) \cos(\alpha) \sin(\kappa t) - X(\alpha) \sin(\alpha) (1 - \cos(\kappa t)), \nonumber \\
y(\alpha, t) = && \gamma X(\alpha) \cos(\alpha) (\cos(\kappa t) - 1) - \nonumber \\ 
&& \gamma X(\alpha) \sin(\alpha) \bigg\{\sin(\kappa t) + \frac{t}{\gamma} \left( \frac{\kappa^2}{2 \omega} - 2 \omega \right) \bigg\},
\label{ePosition}
\end{eqnarray}
\noindent
where $\omega$ and $\kappa$ is the orbital and epicyclic frequency, respectively, and $\gamma = 2\omega/\kappa$ \citep{Binney2008}.

The value of the epicyclic semi-minor axis $X$ depends not only on the velocity $\widetilde{v}_{\rm e,I}$, but also 
on the direction of escape $\alpha$. 
From eqs. (9) and (10) of paper I, it follows
\begin{equation}
X(\alpha) = \frac{\gamma \widetilde{v}_{\rm e,I}}{\kappa} \frac{1}{\sqrt{\gamma^2 \cos^2 (\alpha) + \sin^2 (\alpha)}}.
\label{eXasAlpha}
\end{equation}

\subsection{The tilt of the tidal tail}

At time $t$, stars which escaped at velocity $\widetilde{v}_{\rm e,I}$ are located 
at the curve given by \eq{ePosition} with $X$ substituted from \eq{eXasAlpha}. 
\reff{ftailShape} shows the shape of the tail I formed by stars escaping at $\widetilde{v}_{\rm e,I} = 2 \Kms$ at seven 
time events for the frequencies $\omega = 8.381 \times 10^{-16} \Si$ and 
$\kappa = 1.185 \times 10^{-15} \Si$, which are expected at the Solar circle (at Galactocentric radius $R_{\rm 0} = 8.5 \Kpc$) 
for the Galaxy model by \citet{Allen1991}. 

At the age of $40 \Myr$, the tail points almost towards the Galactic centre (red contour in \reff{ftailShape}), and with 
progressing time the tail aligns with the direction of the orbit around the Galaxy (at $160 \Myr$; blue contour). 
Later, the tail is again tilted with respect to its orbit ($240 \Myr$; yellow contour), 
and these oscillations continue with time.  

At a given time $t$, the tail is parametrised by the angle $\alpha$. 
Thus, a quantitative description of the tail tilt can be obtained by finding the value of $\alpha$ for which the distance 
from the cluster $r = \sqrt{x^2 + y^2}$ is the largest. 
The value of $\alpha_{\rm ext}$, which extremalises the distance on the tail can be found by setting $\pder{r}{\alpha} = 0$, which results in 
\begin{eqnarray}
&& \frac{(\widetilde{x}^2 + \widetilde{y}^2)(\gamma^2 - 1)}{\gamma^2 \cos^2 (\alpha_{\rm ext}) + \sin^2 (\alpha_{\rm ext})} (\sin(\alpha_{\rm ext}) \cos(\alpha_{\rm ext})) - \nonumber \\
&&  \widetilde{x} \left(\sin(\alpha_{\rm ext}) \sin(\kappa t) + \cos(\alpha_{\rm ext})(1 - \cos(\kappa t)) \right) + \nonumber \\
&& \gamma \widetilde{y} \Bigg(\sin(\alpha_{\rm ext})(1 - \cos(\kappa t)) - \nonumber \\
&& \cos(\alpha_{\rm ext}) \left\{\sin(\kappa t) + \frac{t}{\gamma}\frac{(\kappa^2 - 4 \omega^2)}{2 \omega} \right\} \Bigg) = 0,
\label{eAlphaExt}
\end{eqnarray}
where the dimensionless quantities $\widetilde{x}$ and $\widetilde{y}$ are defined as $\widetilde{x} = x/X$ and $\widetilde{y} = y/X$, respectively.

\iffigs
\begin{figure}
\includegraphics[width=\columnwidth]{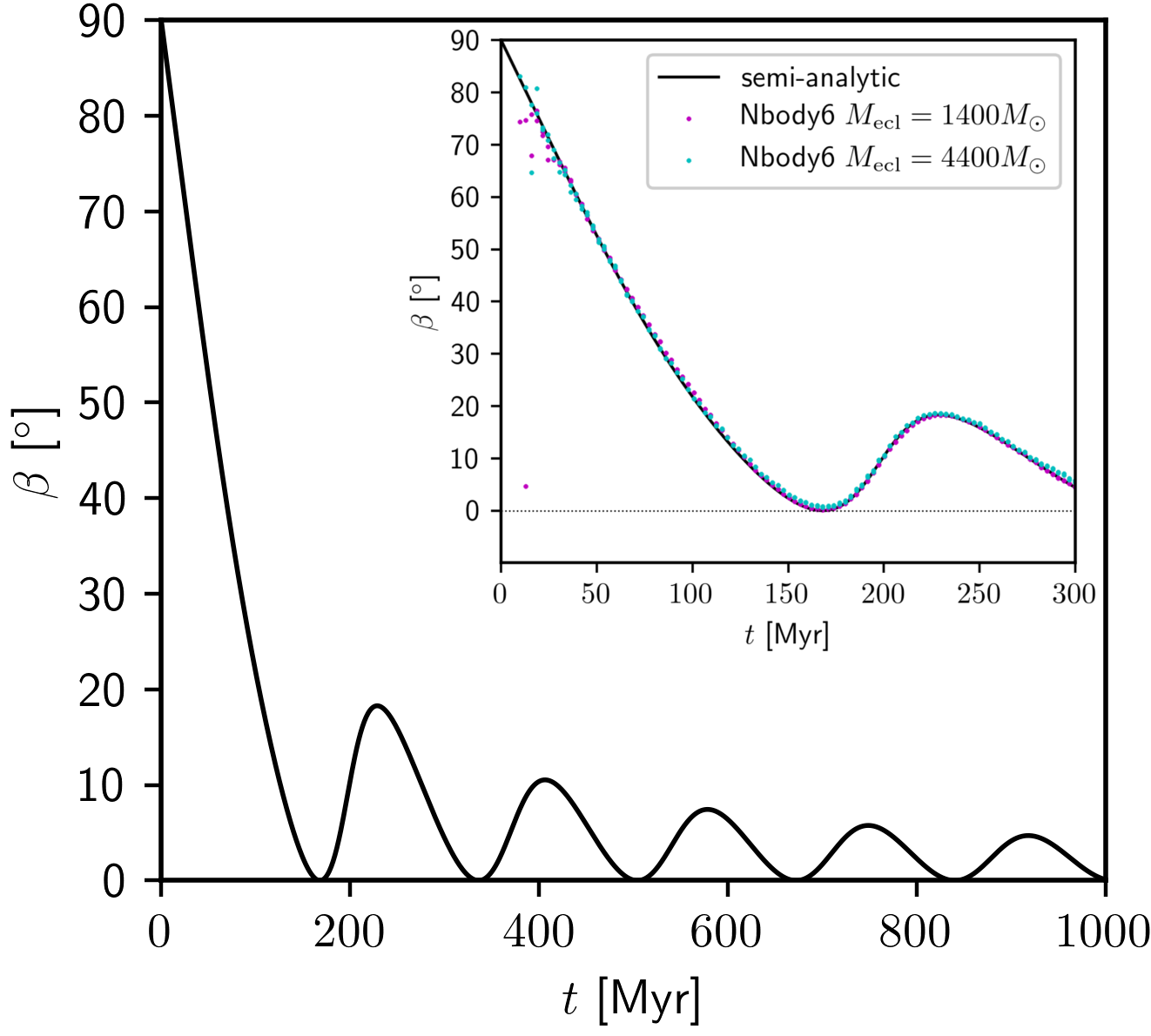}
\caption{The time dependence of the theoretical tail tilt angle $\beta$ (according to eqs. \ref{eAlphaExt} and \ref{eBeta}). 
The value of $\beta$ at a given age is independent of the velocity of escaping stars $\widetilde{v}_{\rm e,I}$ or any other cluster property. 
Inset: Comparison of the theoretical solution (black line) with N-body models of initial stellar mass
$M_{\rm ecl} = 1400 \Msun$ and $M_{\rm ecl} = 4400 \Msun$ (dots). 
}
\label{fTimeEvolv}
\end{figure} \else \fi

In the interval $\alpha \in (-\pi/2, \pi/2)$, \eq{eAlphaExt} has typically two solutions: one corresponding to the maximum distance $r$ (at angle $\alpha_{\rm ext}^{+}$), 
and the other corresponding to the minimum distance $r$ (at angle $\alpha_{\rm ext}^{-}$). 
The tilt of the tail is the angle $\beta$ at which the tip of the tail is seen from the cluster relatively to the positive $y$-axis (\reff{ftailShape}), i.e. 
\begin{equation}
\tan(\beta) = -\frac{x(\alpha_{\rm ext}^{+}(t), t)}{y(\alpha_{\rm ext}^{+}(t), t)}.
\label{eBeta}
\end{equation}
At a given $t$, the angles $\alpha_{\rm ext}^{+}$ and $\alpha_{\rm ext}^{-}$ depend only on the Galactic frequencies $\omega$ and $\kappa$, and they 
are independent on any property of the cluster such as for example $\widetilde{v}_{\rm e,I}$. 
Thus, the tidal structure occupied by stars escaping the cluster at different velocity $\widetilde{v}_{\rm e,I}$ 
forms concentric curves with respect to the origin; a group of stars escaping at velocity $k \times$ larger reaches a $k \times$ larger distance from the cluster 
in any direction in the $x$-$y$ plane (eq. \ref{eXasAlpha}), but the tidal structure is concentric with the same tilt for 
a group of stars of any $\widetilde{v}_{\rm e,I}$.
The time evolution of the angle $\beta$ is shown in \reff{fTimeEvolv}. 

To check our calculations, we compared the value of $\beta$ obtained from eqs. \ref{eAlphaExt} 
and  \ref{eBeta} with \nbdvi models of $M_{\rm ecl} = 1400 \Msun$ and $4400 \Msun$ clusters 
with rapid gas expulsion and a star formation efficiency (SFE) of 33\%, 
which were studied in paper II (see there the description of models C03G13 and C10G13 for details). 
In the numerical models, we calculate the tilt as the angle of the eigenvector corresponding to the long axis of the covariance matrix from the stellar 
$x$ and $y$ positions projected to the Galactic mid-plane.
The comparison, which is shown in the inset of \reff{fTimeEvolv}, demonstrates very 
good agreement with the semi-analytical formulation (eqs. \ref{eAlphaExt} and \ref{eBeta}) for most of the time apart from the youngest age. 
The scatter at the youngest age ($t \lesssim 20 \Myr$) is caused by the difficulty of finding the semi-major axis of the tail, which 
is of an almost spherical shape at this time.

Although the dating method uses stellar kinematics, because it is based on \eq{ePosition}, 
the only important observational quantity is the tail tilt angle $\beta$. 
This might make the method simpler to use than standard kinematic methods, because it does not require stellar velocities for the age determination, 
velocities being used only to identify the tidal tail.
For this reason, we will refer in the following to this method as to the \textit{morphological method} (and corresponding ages as \textit{morphological ages}).
The standard methods of cluster age determination, which are based on stellar evolution (e.g. isochrone fitting, lithium depletion boundary and gyrochronology), 
are refereed to as stellar evolutionary methods (and corresponding ages as stellar evolutionary ages).
Another advantage of the present method is its insensitivity to completeness because $\beta$
can be estimated only from a fraction of the stars in the tail.

\subsection{The width and length of the tidal tail}

\iffigs
\begin{figure}
\includegraphics[width=\columnwidth]{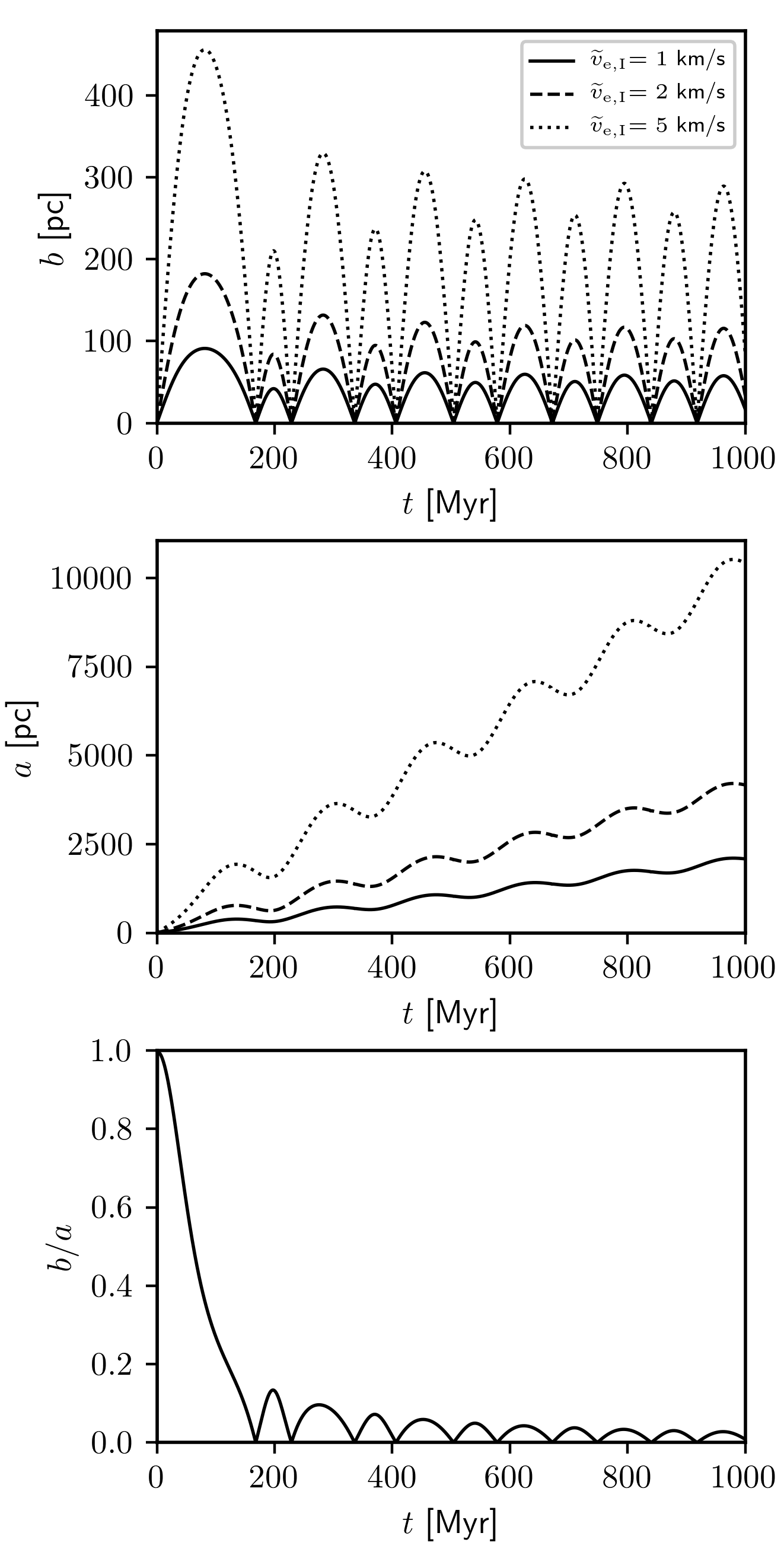}
\caption{
The time dependence of the width of the tidal tail (top panel), 
the tail length along its longest axis (middle panel), 
and the tail aspect ratio (lower panel). 
The different line styles in the top and middle panels represent expansion speeds $\widetilde{v}_{\rm e,I}$ of $1$, $2$ and $5 \Kms$.
}
\label{fExtent}
\end{figure} \else \fi

The width $b$ of the tail is attained at the solution to \eq{eAlphaExt} which corresponds to the minimum distance, 
i.e. at angle $\alpha_{\rm ext}^{-}$,
\begin{equation}
b(t) = 2 \sqrt{x^2(\alpha_{\rm ext}^{-}(t), t) + y^2(\alpha_{\rm ext}^{-}(t), t)}. 
\label{etThickness}
\end{equation}
At a given age, $b(t)$ is a linear function of $\widetilde{v}_{\rm e,I}$ (eq. \ref{eXasAlpha}). 
The time dependence of $b$ for $\widetilde{v}_{\rm e,I} = 1 \Kms$, $2 \Kms$ and $5 \Kms$ is shown in the upper panel of \reff{fExtent}.
We note that although the shape of tail I resembles an ellipse, the curve is not an exact ellipse, because \eq{ePosition} does not 
represent a parametric equation of an ellipse.

Likewise, the tail reaches its longest distance $a$ from the origin for stars escaping at the angle $\alpha_{\rm ext}^{+}$, 
\begin{equation}
a(t) = 2 \sqrt{x^2(\alpha_{\rm ext}^{+}(t), t) + y^2(\alpha_{\rm ext}^{+}(t), t)}.
\label{etLength}
\end{equation}
The tail length obtained by this way for the three values of $\widetilde{v}_{\rm e,I}$ is shown in the middle panel of \reff{fExtent}.
The lower panel of the figure plots the time evolution of the aspect ratio of the tail, $b(t)/a(t)$, which shows 
that the tail gets more elongated with time.

\section{Application to the star clusters with known extended tails and unbound stellar streams}

\label{sApplication}

\subsection{Age determination}

\label{ssDynAge}

\iffigs
\begin{figure}
\includegraphics[width=\columnwidth]{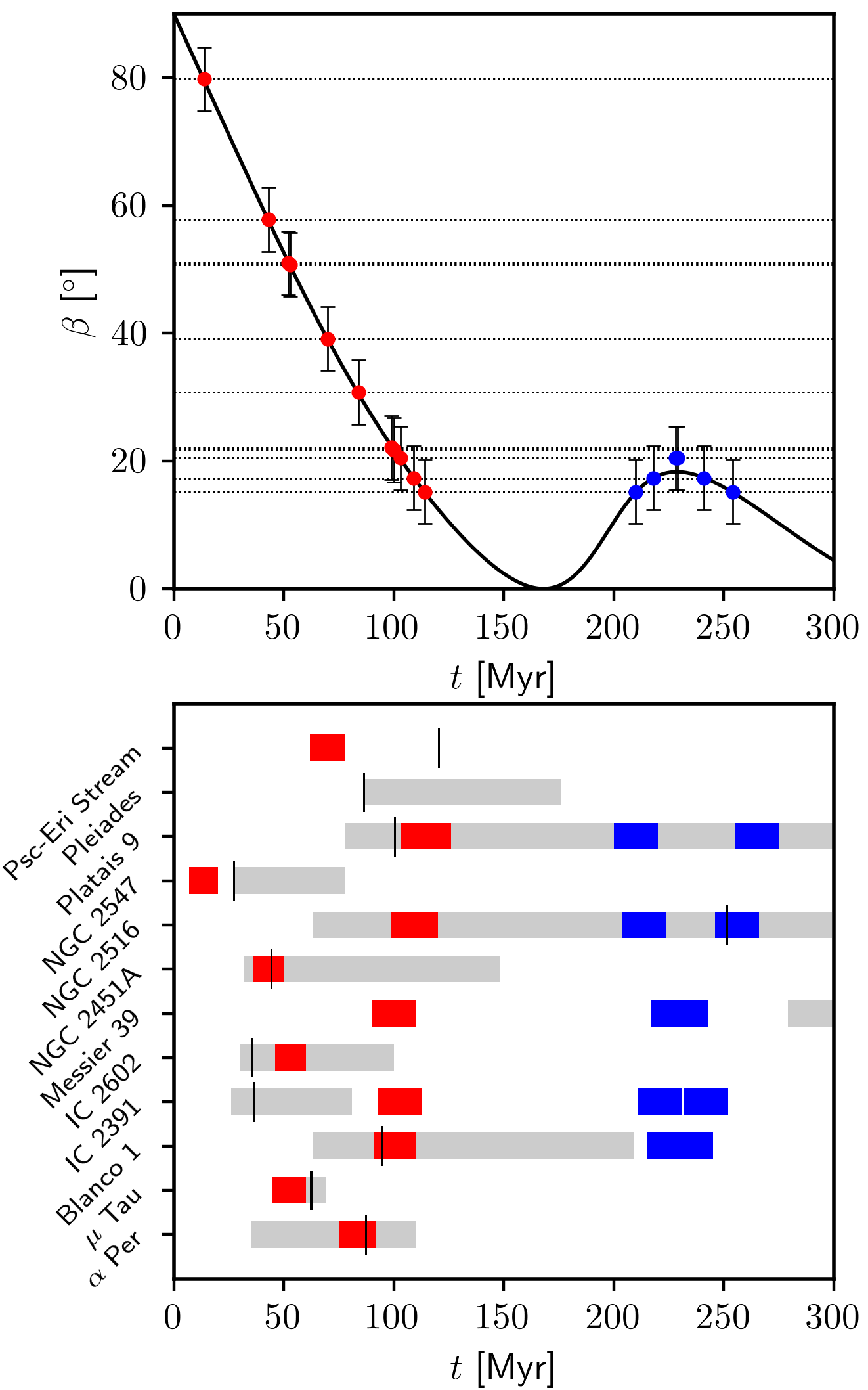}
\caption{
The morphological age determination for the star clusters with known extended tail structures from \citet{Meingast2021}, and for the 
Psc-Eri and $\mu$ Tauri streams \citep{Meingast2019,Gagne2020}. 
The upper panel shows the ages which correspond to the observed values of $\beta$. 
The degeneracy of age for a given $\beta$ is illustrated by the dot colour: the age corresponding to the first and second tail 
oscillation is shown by the red and blue dots, respectively.
The typical uncertainty in measuring $\beta$ is assumed to be $5 ^{\circ}$ (error bars). 
\figpan{Lower panel:} Comparison between the morphological cluster age determination from the tail tilt (red and blue bars) 
with the age estimates based on stellar evolution (grey bars; see \reft{tageEst} for details).
The most probable cluster age $t_{\rm sev}$ is indicated by the black vertical bars.
}
\label{fCompObs}
\end{figure} \else \fi

\begin{table*}
\begin{tabular}{lcccc|cccc}
Object name & $t_{\rm sev, min}$ & $t_{\rm sev, max}$ & $t_{\rm sev}$ & $M_{\rm ecl, obs}$ & $\beta$ & $t_{\rm mph, min}$ & $t_{\rm mph, max}$ & $M_{\rm ecl, mph}$  \\
& [Myr] & [Myr] & [Myr] & [$\rm{M}_{\odot}$] & [$^\circ$] & [Myr] & [Myr] & [$\rm{M}_{\odot}$] \\
\hline

      Platais 9 &     78 &    347 &    100 &  285 &   15 &    103 &    126 & 20 \\ 
     Messier 39 &    279 &   1023 &    310 &  325 &   22 &    215 &    245 & 20 \\ 
      $\alpha$ Per &     35 &    110 &     87 & 1030 &   31 &     75 &     92 & 60 \\ 
      NGC 2451A &     32 &    148 &     44 &  425 &   58 &     36 &     50 & 20 \\ 
        IC 2602 &     30 &    100 &     35 &  400 &   51 &     46 &     60 & 20 \\ 
       NGC 2547 &     27 &     78 &     27 &  590 &   80 &      7 &     20 & 870 \\ 
       Blanco 1 &     63 &    209 &     94 &  365 &   22 &     91 &    110 & 5 \\ 
        IC 2391 &     26 &     81 &     36 &  445 &   21 &     93 &    113 & 50 \\ 
       NGC 2516 &     63 &    299 &    251 & 2550 &   17 &    205 &    265 & 30 \\ 
      Pleiades &     86 &    176 &     86 &   850 &  -32 &  - &  - & - \\ 
\hline
  Psc-Eri Stream &    120 &    120 &    120 & $\gtrsim 2000$  &    39 &     62 &     78 & 200  \\ 
         $\mu$ Tau &     55 &     69 &     62 & $\approx 250$ &   51 &     45 &     60 & 1200
\end{tabular}
\caption{Morphological age and mass estimates for ten open star clusters and two tidal streams (Psc-Eri stream and $\mu$ Tau). 
The minimum, maximum and the most probable stellar evolutionary age estimate is denoted by $t_{\rm sev, min}$, $t_{\rm sev, max}$ and $t_{\rm sev}$, respectively. 
The measured angle of the tail tilt and the minimum and maximum morphological age is denoted $\beta$, $t_{\rm mph, min}$ and $t_{\rm mph, max}$, respectively; 
and the observed and expected mass is denoted by $M_{\rm ecl, obs}$ and $M_{\rm ecl, mph}$, respectively.
For Messier~39 and NGC~2516, we give the morphological age estimates during the second tail oscillation as these are in a better agreement with the 
stellar evolutionary ages. 
The stellar evolutionary age and mass estimates are adopted from \citet{Meingast2021} for the open clusters (above the horizontal line), 
and from \citet{Curtis2019}, \citet{Ratzenbock2020} and \citet{Gagne2020} for the Psc-Eri and $\mu$ Tau streams. 
}
\label{tageEst}
\end{table*}

We illustrate the proposed age determination method on data which we compiled from the existing literature. 
The examples below serve only as a consistency check of the method, and they are not meant for improvement of the current age estimates of 
these clusters and streams because the angle $\beta$ was estimated only by eye from $x$-$y$ maps of the tidal structures, and 
we assume that the clusters or streams orbit the Galaxy on exactly circular trajectories (see \refs{ssSevComp} for more details).

We took as the basis for our cluster sample the $x$-$y$ maps from \citet{Meingast2021} (their figure A.2). 
In addition to these structures, which still surround gravitationally bound clusters, we include two tidal streams into our sample: 
the Psc-Eri stream from \citet{Meingast2019} with an age determination of \citet{Curtis2019}, and the $\mu$ Tauri association from \citet{Gagne2020}. 

\reft{tageEst} lists the measured tilt angle $\beta$, and the estimated age range ($t_{\rm mph,min}$, $t_{\rm mph,max}$) 
obtained by inverting the function $\beta = \beta(t)$ from \eq{eBeta} as illustrated in the top panel of \reff{fCompObs}. 
We estimate that the angle $\beta$ is measured with an uncertainty of $\Delta \beta = \pm 5^\circ$ for each object
\footnote{The estimate is based on the assumption that the tidal tail is divided into $k_{\rm seg}$ equal azimuthal segments 
with the origin at the cluster, 
and that the number of stars $n_{\rm seg}$ in each segment is distributed according to the Poisson distribution. 
Requiring the ratio between the mean and standard deviation as $\langle n_{\rm seg} \rangle / \sigma_{\rm n_{\rm seg}} \gtrsim  2$, 
we obtain $\langle n_{\rm seg} \rangle \gtrsim 4$. 
Further assuming that each tidal tail contains typically at least $300$ stars \citep[][their fig. 12]{Meingast2021}, 
one obtains $k_{\rm seg} = 300/\langle n_{\rm seg} \rangle = 75$ azimuthal segments, i.e. each segment spans $4.8 ^\circ$, which we identify with $\Delta \beta$. 
This is an order of magnitude estimate as the signal to noise ratio of $2$ and the number of stars in the tail of $\gtrsim 300$ might be too optimistic. 
On the other hand, the estimate is provided for a non-elongated tidal tail; an elongated tidal tail has more stars in the segments 
along its longer axis, which increases the signal to noise ratio. 
}
, 
and the uncertainty is propagated to the age uncertainty. 
The red and blue points represent, respectively, the age estimate for the first (up to $168 \Myr$) 
and second (between $168 \Myr$ and $336 \Myr$) oscillation of the tail, and they 
illustrate that the age determination is degenerated for objects with $\beta - \Delta \beta < 18 ^\circ$ (Platais 9, NGC~2516, Messier~39, IC~2391 and Blanco~1).  
Because of the degeneracy, it is useful to have a-priory age knowledge for objects with $\beta - \Delta \beta < 18 ^\circ$ so that the 
age can be searched for either during the first or second oscillation.
The red and blue bars in the lower panel of \reff{fCompObs} represent the estimated morphological age interval ($t_{\rm mph,min}$, $t_{\rm mph,max}$) 
during the first and second tail oscillation, respectively. 
The range of age determination from various stellar evolutionary signposts is indicated by the grey bars, 
and the most probable age (according to \citealt{Meingast2021}) is indicated by the black bars.

We exclude the Pleiades from further analysis because there has been no prominent tidal tail found around the Pleiades so far, and the putative tail 
found in the data of \citet{Meingast2021} points in the direction to the Galactic anti-centre (i.e. $\beta < 0$), for which no age solution exists. 
From the $11$ objects left, six (Platais~9, NGC~2516, NGC~2451A, Blanco~1, $\mu$ Tau and $\alpha$ Per) have the 
morphological age range either in complete agreement with or differing at maximum by $10\%$ 
from the most probable stellar evolutionary age. 
A notable example is NGC~2516, which agrees with its stellar evolutionary age during its second tail tilt (not the first tilt), which 
might constrain its age to the interval from $\approx 200$ to $260 \Myr$
\footnote{Using gyrochronology, \citet{Bouma2021} find a slightly younger age of $\approx 150 \Myr$.}
.

Three other clusters (NGC~2547, Messier~39, and IC~2602) differ more from their stellar evolutionary ages, but the difference is not huge (\reff{fCompObs}): 
In the case of NGC~2547, $t_{\rm sev}$ differs from the lower morphological estimate $t_{\rm mph,min}$ by only $7 \Myr$ (see \reft{tageEst} for details); 
For Messier~39, $t_{\rm sev} = 310 \Myr$ differs by a factor of $1.3$ from $t_{\rm mph, max} = 245 \Myr$;
and the morphological age estimate of IC~2602 lies within the interval allowed by the stellar evolutionary age. 
Only two objects have $t_{\rm sev}$ substantially different from the morphological age estimate (Psc-Eri stream and IC~2391).

\subsection{Estimate of the initial embedded mass}

\label{ssMass}

\iffigs
\begin{figure*}
\includegraphics[width=\textwidth]{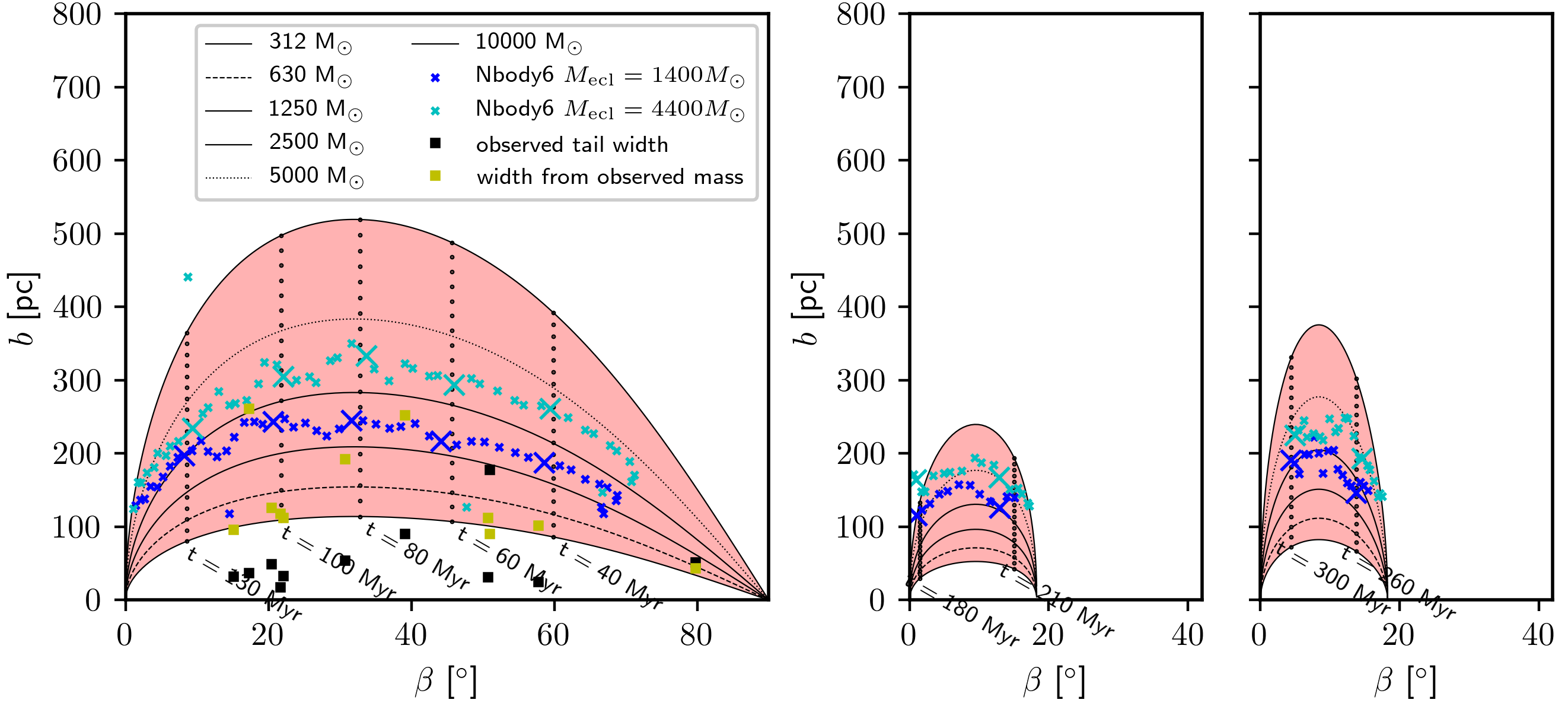}
\caption{The relationship between the angle $\beta$ and the tail width $b$ as a function of the cluster age (black dots at selected times), 
and the initial cluster mass $M_{\rm ecl}$ (lines). 
The area enclosing clusters of mass from $312 \Msun$ to $10^4 \Msun$ is filled red.
The panels from left to right show the intervals between minima of tail thickness, where $\beta$ evolves monotonically: 
$\beta$ decreases from $0$ to $168 \Myr$ (left panel), increases from $168$ to $228.7 \Myr$ (middle panel), and decreases from $228.7$ to $336.1 \Myr$ (right panel).
For a given $\beta$, the tail width increases monotonically with the cluster mass. 
The blue and cyan crosses represent \nbdvi simulations of the $M_{\rm ecl} = 1400 \Msun$ and $4400 \Msun$ clusters, respectively. 
The simulated point of the age nearest to the selected ages (black dots) is shown by the large cross. 
The width of tails of the observed clusters and streams of \reft{tageEst} are indicated by the black 
squares. 
The expected tail width for the same clusters, but calculated from their observed mass $M_{\rm ecl,obs}$, is indicated by the yellow squares. 
The majority of the observed tails are too narrow for their observed mass. 
}
\label{fAngleThickness}
\end{figure*} \else \fi

The stars forming tidal tail I escape the cluster at a typical velocity $\widetilde{v}_{\rm e,I}$, which is proportional 
to the initial velocity dispersion $\sigma$ in the embedded cluster. 
The velocity dispersion is related to the initial cluster mass $M_{\rm ecl}$ and virial radius $R_{\rm V}$ by $\sigma^2 = G M_{\rm ecl}/(2 R_{\rm V})$ \citep{Aarseth2003}. 
For a cluster represented by the Plummer model of scale radius $a_{\rm Pl}$, for which $R_{\rm V} = 16 a_{\rm Pl} /(3 \pi)$, the velocity dispersion reads
$\sigma^2 = 3 \pi G M_{\rm ecl}/(32 a_{\rm Pl})$.
Assuming that the cluster length-scale $a_{\rm Pl}$ depends only on the cluster mass \citep{Marks2012}, the velocity dispersion can be expressed only as a function of 
the initial (embedded) mass in stars, $M_{\rm ecl}$. 
Thus, from the extent of the tidal tail at a given age, we can determine $\widetilde{v}_{\rm e,I}$ (from eqs. \ref{eXasAlpha} and  \ref{etThickness}), 
and from this quantity $\sigma$ because $\sigma \propto \widetilde{v}_{\rm e,I}$, and finally from $\sigma$ we estimate the initial cluster mass. 
Because we expect the volume density of tail I to be the highest (and the least contaminated) along its short axis, 
we estimate the tail extent (i.e. its width) in this direction. 

We take the typical velocity of tail I stars as $\widetilde{v}_{\rm e,I} = 4 \Kms$ for $M_{\rm ecl} = 4400 \Msun$ (table 1 in paper II), and 
from the arguments above it follows that 
\begin{equation}
\frac{\widetilde{v}_{\rm e,I}}{\Kms} = 5.7 \left(\frac{M_{\rm ecl}}{10^4 \Msun}\right)^{7/16}.
\label{evelMass}
\end{equation}
The tail width $b$ expressed from \eq{etThickness} is shown by lines as a function of $\beta$ in \reff{fAngleThickness}. 
Each of the panel corresponds to a time interval when the tail tilt evolves monotonically: the left panel is for the first 
tail oscillation ($\beta$ decreases), the middle panel for the first part of the second oscillation ($\beta$ increases), 
and the right panel for the second part of the second oscillation ($\beta$ decreases).
For comparison, we plot there the tail width for N-body models C03G13 and C10G13 (blue and cyan crosses, respectively) of paper II, which is 
calculated as the variance of the stellar distances in the tidal tail (i.e. excluding the stars in the cluster) in the direction of the tail semi-minor axis.
The N-body models, which have cluster masses of $1400 \Msun$ and $4400 \Msun$, fit at the expected positions in the $\beta - b$ plane between the 
theoretical curves for $1250 \Msun$, $2500 \Msun$ and $5000 \Msun$ for most of the time.

The tail widths for the observed clusters and streams are plotted by black squares in \reff{fAngleThickness}. 
To provide an order of magnitude estimate, we neglect the uncertainty in the measurements of tail widths, which is probably dominated by incompleteness as 
discussed below.
For simplicity, we show them during the first oscillation only. 
For the tail width at given $t$, we obtain the velocity $\widetilde{v}_{\rm e,I}$ from \eq{etThickness}, and the estimated initial mass $M_{\rm ecl,mph}$
from \eq{evelMass}. 
However, the mass obtained by this way is substantially smaller (often by a factor of $10$; \reft{tageEst}) for most of the objects 
than the observed mass $M_{\rm ecl,obs}$ of the cluster and tail combined. 
In other words, when the expected tail width is calculated from the observed cluster mass $M_{\rm ecl,obs}$, it is substantially (by a factor of $3$) larger 
than its observed value (yellow symbols in \reff{fAngleThickness}) for most of the objects. 
Possible reasons of the discrepancy are discussed in \refs{ssMeclEst}.

\section{Discussion}

\label{sDiscussion}

\subsection{Comparison to stellar evolutionary age estimates}

\label{ssSevComp}

The analysis of \refs{ssDynAge} provides age estimates that are in very good agreement with stellar evolutionary age 
estimates in $\approx 55$ \% of the cases, in reasonable agreement in $\approx 25$ \% of cases, and in tension in $\approx 20$ \% of cases.
The agreement for the majority of cases gives some support to the morphological method. 
The discrepancy in the age for $\approx 20$ \% of cases (the Psc-Eri stream and IC~2391) indicates that one of the methods is incorrect. 
It is possible that it is either one of the stellar evolutionary methods or the morphological method. 
For example, the main sequence fitting method for the 
Psc-Eri stream provided an age of $\approx 1 \Gyr$ \citep{Meingast2019}, which was later found to be younger by a factor of $8$ \citep{Curtis2019}
because of incompleteness in the former data.
An excellent match with the morphological age for both objects could be obtained by changing the stellar evolutionary age by a factor of two, which is 
substantially less than what was the discrepancy for the Psc-Eri stream in the example above.

The morphological method can be further improved by taking into account the non-circularity of the orbit, and the possible influence 
of the Galactic bar. 
The degree of non-circularity (eccentricity) can be obtained from the proper 
motion of the cluster or stream, and then utilised for a correction of the relationship between the morphological 
age and angle $\beta$ for the given eccentricity. 
It is possible that including the effect of the orbital eccentricity would also improve the agreement between the stellar-evolutionary and morphological 
age estimates. 

Giant molecular clouds passing through a tidal tail can stir asymmetry in the tail \citep{Jerabkova2021}, which was indicated in 
the tail (in this case tail II) of the Hyades \citep{Roser2019}. 
According to the results of \citet{Jerabkova2021}, giant molecular clouds might distort also tidal tail I 
so that its tilt $\beta$ would no longer be a clear function of the age of the cluster or stream. 

Another mechanism, which is likely to distort the shape of the tail I is gas expulsion occurring off the cluster centre 
(e.g. caused by a massive star, which was sent to the outskirts of the still embedded cluster by an encounter near the centre).  
This would accelerate stars in one direction at a larger velocity than in the opposite direction, breaking the assumption of isotropy.
We intend to explore some of these possibilities in a follow up paper.

\subsection{Applications and limitations}

\label{ssLimitations}

Morphological age determinations can be applied to both tidal tails enveloping gravitationally bound clusters 
from which the tails presumably originated as well as to stellar streams, which contain no gravitationally bound remnant cluster. 
Moreover, it is not necessary that the stellar stream originated from a gravitationally bound object. 
For example, consider a star forming region with sparse star formation occurring only in small groups or clusters,
which completely dissolve and release all their stars to the field during or soon after the star forming cloud is disrupted. 
These stars will expand in random directions from the location of their birth clusters, and will be subjected to the Galactic tidal field. 
At a given time $t$, stars released from each cluster will occupy an area approximately bounded by the contour of \reff{ftailShape} corresponding to $t$ 
and scaled by the typical escape velocity $\widetilde{v}_{\rm e,I}$ from each individual cluster. 

Whether the morphological dating provides robust estimate depends mainly on the volume occupied by the star forming region. 
Since the majority of star formation in the Galaxy occur in filaments \citep{Andre2014}, the stars which formed in the filaments 
occupy a filamentary configuration long after the star forming region became inactive with all gas removed \citep{Jerabkova2019}.  
Stellar relic filaments reach sizes up to several hundreds of parsec \citep{Jerabkova2019,Kounkel2019,Beccari2020,Wang2021}, and often 
span large distances between gravitationally bound clusters \citep{Beccari2020}. 
In this case, parts of relic filaments might be confused with tidal tails, which probably posses the largest limitation for the morphological dating method.

On the other hand, many star-forming regions occupy smaller volumes 
(e.g. the maximum extent of the Taurus-Auriga star-forming region is $\approx 60 \Pc$; \citealt{Galli2019}) 
than the volume of the tail already at a young age ($\approx 100 \Pc$; \reff{ftailShape}), the tidal tails from the clusters will be 
superimposed one over another, overlapping and probably being indistinguishable from each other. 
Nevertheless, the tilt $\beta$ of the tail will be comparable for all the clusters (the age spread of clusters forming within the same 
star-forming region is $\lesssim 25 \Myr$; e.g. \citealt{Kawamura2009,Jeffries2011,Dobbs2013}, which is much less than $2 \pi/\kappa \approx 170 \Myr$), 
so the morphological dating method can be applied to initially unbound configurations as well. 

In order to estimate the limitations due to relic filaments, we consider that a typical relic filament is of length of $100 \Pc$, 
and that the filament forms a star cluster at one of its extremities. 
The cluster releases stars to tail I at typical speeds of $\widetilde{v}_{\rm e,I} \approx 2 - 3 \Kms$. 
These stars will overtake the relic filament at an age of $\approx 30 - 50 \Myr$, 
when both the filament and the tidal tail will be clearly discernible from one another. 
This time is also near the upper limit on the observed age of relic filaments \citep{Jerabkova2019,Beccari2020}, and the age when 
the elongation of the tail I becomes less spherical so that its tilt can be measured (see the red contour in \reff{ftailShape}). 
At the same time, the density of the relic filament decreases with time as the filament expands radially, lowering its contamination of tail I.
Thus, if the cluster forms as a part of a large filamentary structure, its morphological age can be obtained after it is older than $\approx 40 \Myr$.

Another complication in determining the morphological age are the stars which evaporate from the cluster and form tail II. 
These stars occupy a very similar area in the position-velocity space as stars from tail I so that both tails would be probably indistinguishable in the Gaia data. 
This impacts in particular older clusters because stars forming tail II evaporate from clusters 
at an approximately time-independent rate \citep{Baumgardt2003,Aarseth2003,Chumak2006a}. 
Also, stars in tail II move at lower speeds than stars in tail I (Paper II), so tail II is to be found closer to the cluster and at 
an elevated stellar density. 
Unlike tail I, tail II is S-shaped in the vicinity of the cluster \citep{Chumak2006a,Kupper2008}, which would spuriously increase $\beta$ 
whereupon underestimating the proper cluster age. 

In order to suppress contamination by tail II, we suggest that for measuring $\beta$, only sufficiently distant stars from the cluster are 
taken into account. 
Since the typical speeds in tail II are relatively low ($\widetilde{v}_{\rm e,II} \approx 1.4 \Kms$ for a rather massive cluster of $4400 \Msun$ and 
lower for lower mass clusters; see table 1 of Paper II), the analysis should ignore stars located closer than $\approx \widetilde{v}_{\rm e,II} t$ 
to the cluster, where $t$ is the prior estimate of the cluster age. 

How many objects do we expect to be accessible for the method given the astrometry of the Gaia DR2 release?
The velocity cut of $2 \Kms$ can be provided for A0 stars closer than $1 \Kpc$ \citep{Gaiac2016b}. 
At this distance, the position error is $\lesssim 10 \Pc$, which should be sufficient for the detection of tail I. 
\citet{Porras2003} find $16$ very young star clusters having more than $100$ stars within the circle of radius $1 \Kpc$ centred at the Sun. 
For this estimate, we require that only more numerous clusters (having more than $800$ stars) can produce tails of discernible tilt. 
Assuming that the Galaxy forms embedded clusters according to an embedded cluster mass function in the form of 
$\nderrow{N_{\rm ecl}(M_{\rm ecl})}{M_{\rm ecl}} \propto M_{\rm ecl}^{-2}$ \citep[e.g.][]{Lada2003,FuenteMarcos2004}, 
spanning up to $M_{\rm ecl} \approx 10^4 \Msun$ \citep{Johnson2017}, 
$5/8$ of the star clusters with more than $100$ stars contain more than $800$ stars. 
Further assuming that the catalogue of \citet{Porras2003} is complete for clusters younger than $5 \Myr$ and that the morphological 
method can be used for clusters in the age interval from $40$ to $340 \Myr$ (see \refs{ssAgeRange}), 
we obtain $5/8 \times 16 \times (340 - 40)/5 \approx 550$ objects (i.e. clusters or streams) 
accessible to the method within $1 \Kpc$ from the Sun.
Even though this is the likely upper estimate (because many of the clusters form within the same star-forming region so that their tails cannot be 
distinguished from each other and the upper mass of clusters which form within the Solar circle is probably lower than $10^4 \Msun$ 
according to \citealt{PflammAltenburg2008}), 
we estimate that there might be more than one hundred objects in the Gaia DR2 data accessible for this dating method. 

\subsection{The range of cluster ages available for the morphological dating method and its accuracy}

\label{ssAgeRange}

The morphological age determination method is suited for younger clusters, where $\beta$ sensitively depends on the age. 
This is mainly because the function $\beta = \beta(t)$ is not an injective function; there is one $t$ for $\beta \gtrsim 18.3 ^\circ$, three possible values 
for $t$ for $\beta \in (10.5 ^\circ, 18.3 ^\circ)$, five possible values of $t$ for $\beta \in (7.4 ^\circ, 10.5 ^\circ)$, and the degeneracy quickly 
increases for $\beta \lesssim 7.4 ^\circ$ (\reff{fTimeEvolv}). 
Any systematic error in the determination of $\beta$ (e.g. from the cluster having an eccentric orbit or confusion between tail I and tail II stars), results in 
a large uncertainty in age $t$ for $\beta$ sufficiently small. 
Taking $\Delta \beta = 5 ^\circ$ as an estimate of the uncertainty in $\beta$ determination, this method is useless for $t \gtrsim 336 \Myr$ 
because $\beta$ changes by less than $\approx 2 \Delta \beta$ since this age. 
On the other hand, $\beta$ changes rapidly for younger clusters (it decreases from $\approx 90 ^\circ$ to the smallest angle of uniquely 
determined time of $18.3 ^\circ$ in $105 \Myr$) offering a very sensitive tool for clusters in this age range. 
Likewise, $\beta$ is sensitive to $t$ also during the second tail oscillation (between $168 \Myr$ and $336 \Myr$). 
The knowledge of a prior estimate of the tail age (e.g. from a stellar evolutionary method) would be useful because it would 
constrain whether the tail is in its first or second oscillation, 
and then the inversion $t = t(\beta)$ could be done in the appropriate time interval, which would provide the morphological estimate for the age.
We show an example of using the prior estimate for the tail age in \refs{ssDynAge} (for NGC~2516; see also \reff{fCompObs}). 

The form of $\beta = \beta(t)$ and the assumed error of $\Delta \beta = 5 ^\circ$ can be used for estimating the age of clusters younger than $\approx 340 \Myr$. 
On the other hand, the method appears to be less accurate or problematic for clusters younger than $\approx 40 \Myr$ because of the possible presence of relic filaments and 
the initially spherical expansion of tail I (see \refs{ssLimitations} for details). 

The assumed uncertainty in $\beta$ of $\pm 5^\circ$ propagates to an age uncertainty of $\approx 12 \Myr$ during the first tail oscillation (i.e. at ages younger 
than $168 \Myr$), and to an age uncertainty of $\approx 30 \Myr$ during the second tail oscillation (i.e. between $168$ and $336 \Myr$). 
This uncertainty, which is of $\approx 10$ to $20$\%, is comparable to the most accurate stellar evolutionary dating methods, i.e. 
isochrone fitting and lithium depletion boundary \citep{Meynet2009,Soderblom2010,Jeffries2013,Martin2018,Binks2021}, 
but the morphological method is much easier to apply to a particular cluster or stream as it does not need stellar evolutionary models or 
dedicated spectroscopic observations. 
However, the role of possible systematic errors (e.g. non-circular orbits or contamination) remains to be clarified.

\subsection{Estimate of the initial cluster mass}

\label{ssMeclEst}

Although the morphological age estimate is in agreement with other dating methods for the majority of the clusters in our sample, the 
estimate of the initial mass (\refs{ssMass}) provides substantially lower cluster masses than observed in most of the cases.
The discrepancy in the initial cluster masses (or equivalently the tail widths) might point to incompleteness in the 
observational data, or possibly contamination with stars of tail II, which is most conspicuous close to the cluster. 
The estimate of cluster mass is based on more assumptions (e.g. on the initial cluster radius and the cluster's virial state) 
than the estimate for the age, which might amplify the uncertainty.
For the order-of-magnitude estimate in this work, we inspected the available data by eye only. 
It is possible that more sophisticated tools (e.g. a Bayesian analysis of the probability that a star belongs to tail I) 
might lead to a substantial improvement of the present estimate.

Alternatively, the same outcome could be caused if the velocity $\widetilde{v}_{\rm e,I}$ is lower than assumed, for example if the clusters form with 
the SFE being larger than $1/3$; or if the gas expulsion time-scale is adiabatic; or if the majority of the extended structure originates from 
stars which were never gravitationally bound to the cluster, but which formed nearby to the cluster in the same star forming region.

\section{Summary}

\label{sSummary}

We propose a new method (morphological method) for dating open star clusters with extended tidal tails 
and stellar streams based on the tilt angle $\beta$ of their extended tidal structure measured from the
direction of their orbit around the Galaxy. 
The tidal tail, which is coeval with the cluster, forms at an early age either from stars released due to expulsion of non-star forming gas from the cluster or 
from stars formed in the same star forming region in the vicinity of the cluster. 
Classical tidal tails forming by gradual evaporation of stars from the cluster cannot be used by this method. 

We show analytically that the tilt angle $\beta$ for objects (i.e. clusters or streams) at a given Galactocentric distance is only a function of the object age $t$ 
and not a function of any other property of the object such as its initial mass or radius. 
The age can be found by inverting the theoretical dependence $\beta = \beta(t)$ for observed $\beta$ (upper panel of \reff{fCompObs}). 
The method is suitable for younger objects ($40 \Myr \lesssim$ age $\lesssim 350 \Myr$), 
where we estimate the accuracy to be $10$ to $20$\% of the age of the object (for an error in $\beta$ of $5 ^\circ$), 
which is comparable to the errors of the currently most accurate stellar evolutionary dating methods, 
which utilise isochrone fitting or the lithium depletion boundary. 
The morphological method does not necessarily aim at exceeding the accuracy of the stellar evolutionary methods, 
but at providing an estimate, which is completely independent of models of stellar evolution.

The main advantage of this method is its ease of use and its independence from 
stellar evolutionary models and stellar velocity measurements (apart from the velocity cut to detect the tidal structure). 
The present derivation applies to clusters or streams on circular orbits only, but its extension to mildly eccentric orbits should be straightforward. 
The orbital eccentricity can be estimated from stellar velocities and then used to correct the results. 
We plan to derive the solution for non-circular orbits in a follow up study.
Possible sources of systematic errors originate from the time-scale and isotropy of gas expulsion, 
contamination due to relic stellar filaments \citep{Jerabkova2019,Beccari2020}, and 
contamination due to the stars which gradually evaporate from the cluster forming the classical tidal tail.

We attempted to use the same analytical approach for estimating the initial mass of the clusters from the width of the tidal structure, 
but this yielded substantially lower masses than the masses of the observed structures for most of the objects. 
We attribute the difference to the sensitivity of the method on the initial conditions before and during gas expulsion or to incompleteness 
of current observational data.

We illustrate morphological dating on a sample of ten open clusters and two unbound stellar streams, 
which we compiled from the literature (\refs{ssDynAge}), and we compare 
the results with standard dating methods based on stellar evolution.
Although we assumed that the clusters and streams orbit the Galaxy on circular trajectories and the stellar evolutionary methods often admit large uncertainties, 
the morphological method agrees very well with stellar evolutionary methods in $55$\% of cases, approximately in $25$\% 
of cases and is in tension in only $20$\% of cases. 
Although the results are encouraging, we caution that the present model is idealised, and future theoretical work is needed for the 
method to become more accurate and for its limitations to be better understood.
Because of the aforementioned limitations, we use these examples only as an illustration of the method 
without attempting to improve the current estimates of the age of the objects.

\begin{acknowledgements}
We would like to thank an anonymous referee for the useful comments, which improved the quality of the paper. 
FD, PK and L\v{S} acknowledge support from the Grant Agency of the Czech Republic under grant number 20-21855S.
TJ acknowledges support from the ESA (European Space Agency) Research Fellowship. 
\end{acknowledgements}


\bibliography{tailTiming}{}
\bibliographystyle{aasjournal}

%
%

\end{document}